\documentclass[pra, aps, twocolumn, superscriptaddress, showpacs, amsmath, amssymb]{revtex4}

\usepackage{graphicx}
\usepackage{float}
\usepackage{verbatim}
\usepackage[usenames,dvipsnames]{color}

\begin{document}

% MACROS
%%%%%%%%%%%%%%%%%%%%%%%%%%%%%%%%%%%%%%%%%%%%%%%%%%%%%%%%%%%%%%%%%%%%%

\newcommand{\beq}{\begin{equation}}
\newcommand{\eeq}{\end{equation}}
\newcommand{\beqa}{\begin{eqnarray}}
\newcommand{\eeqa}{\end{eqnarray}}
\newcommand{\lf}{\hfil \break \break}
\newcommand{\ahat}{\hat{a}}
\newcommand{\adag}{\hat{a}^{\dagger}}
\newcommand{\adagg}{\hat{a}_g^{\dagger}}
\newcommand{\bhat}{\hat{b}}
\newcommand{\bdag}{\hat{b}^{\dagger}}
\newcommand{\bdagg}{\hat{b}_g^{\dagger}}
\newcommand{\chat}{\hat{c}}
\newcommand{\cdag}{\hat{c}^{\dagger}}
\newcommand{\Pihat}{\hat{\Pi}}
\newcommand{\rhohat}{\hat{\rho}}
\newcommand{\shat}{\hat{\sigma}}
\newcommand{\Zhat}{\hat{Z}}
\newcommand{\Xhat}{\hat{X}}
\newcommand{\ket}[1]{\mbox{$|#1\rangle$}}
\newcommand{\bra}[1]{\mbox{$\langle#1|$}}
\newcommand{\ketbra}[2]{\mbox{$|#1\rangle \langle#2|$}}
\newcommand{\braket}[2]{\mbox{$\langle#1|#2\rangle$}}
\newcommand{\bracket}[3]{\mbox{$\langle#1|#2|#3\rangle$}}
\newcommand{\mat}[1]{\overline{\overline{#1}}}
\newcommand{\hak}[1]{\left[ #1 \right]}
\newcommand{\vin}[1]{\langle #1 \rangle}
\newcommand{\abs}[1]{\left| #1 \right|}
\newcommand{\tes}[1]{\left( #1 \right)}
\newcommand{\braces}[1]{\left\{ #1 \right\}}

% HEADER
%%%%%%%%%%%%%%%%%%%%%%%%%%%%%%%%%%%%%%%%%%%%%%%%%%%%%%%%%%%%%%%%%%%%%

\title{Realistic limits on the nonlocality of an \textit{N}-partite single-photon superposition}

\author{Amine Laghaout}
\affiliation{Department of Physics, Technical University of Denmark, Building 309, 2800 Lyngby, Denmark}
\affiliation{Temporary affiliation: NORDITA, Roslagstullsbacken 23, SE-106 91 Stockholm, Sweden}
\author{Gunnar Bj\"{o}rk}
\affiliation{Department of Applied Physics, Royal Institute of Technology (KTH)\\
AlbaNova University Center, SE-106 91 Stockholm, Sweden}
\affiliation{NORDITA, Roslagstullsbacken 23, SE-106 91 Stockholm, Sweden}
\author{Ulrik L. Andersen}
\affiliation{Department of Physics, Technical University of Denmark, Building 309, 2800 Lyngby, Denmark}
\affiliation{NORDITA, Roslagstullsbacken 23, SE-106 91 Stockholm, Sweden}

\date{\today}

\begin{abstract}
A recent paper [L. Heaney, A. Cabello, M. F. Santos, and V. Vedral, New Journal of Physics, \textbf{13}, 053054 (2011)] revealed that a single quantum symmetrically delocalized over $N$ modes, namely a W state, effectively allows for all-versus-nothing proofs of nonlocality in the limit of large $N$. Ideally, this finding opens up the possibility of using the robustness of the W states while realizing the nonlocal behavior previously thought to be exclusive to the more complex class of Greenberger-Horne-Zeilinger (GHZ) states. We show that in practice, however, the slightest decoherence or inefficiency of the Bell measurements on W states will degrade any violation margin gained by scaling to higher $N$. The non-statistical demonstration of nonlocality is thus proved to be impossible in any realistic experiment.
\end{abstract}

\maketitle

% Introduction
%%%%%%%%%%%%%%%%%%%%%%%%%%%%%%%%%%%%%%%%%%%%%%%%%%%%%%%%%%%%%%%%%%%%%
\section{Introduction}

The correlations observed from measuring entangled systems at space-like separated locations may differ depending on whether one adopts a quantum mechanical or a local realistic view of the world. The contradictory predictions of the two perspectives were brought up by Einstein, Podolsky, and Rosen \cite{Einstein} and later enunciated mathematically by Bell in the form of verifiable inequalities binding local realistic correlations to a fixed range \cite{Bell}. This opened up a whole subfield of research where the name of the game is to achieve larger and more conclusive violations of Bell inequalities. The initial formulations of Bell tests were statistical in the sense that the correlations involved had to be evaluated from ensemble measurements. A conceptual breakthrough was achieved when Greenberger, Horne, and Zeilinger (GHZ) proved that a certain class of states could achieve violations in a single run. This came to be known as the ``all-versus-nothing proof'' of nonlocality \cite{Greenberger, Mermin} and was thought to be exclusive to GHZ states, whereby \textit{N} modes populated each by a single particle are superposed to \textit{N} vacuum modes (where $N \geq 3$).

There exists another class of states consisting of a single particle symmetrically distributed over \textit{N} modes, the so-called W states, which can neither be transformed into, nor obtained from, a GHZ state via local operations and classical communication. The most salient difference between the W and the GHZ is that the former are far more robust to noise admixtures and are therefore of practical interest to quantum information protocols \cite{SenDe, Dur, Chaves2010}. It then came as good news that, in addition to being robust to decoherence, the W state also displays an effectively all-versus-nothing nonlocality as the number \textit{N} of delocalizations of the single particle goes up \cite{Heaney}. Given the ease with which W states can be produced, this finding puts them forth as promising candidates in the quest to close the detection loophole that has plagued Bell tests, particularly in the optical regime. We will show in this article, however, that the violation of local realism by W states does not scale as hoped for with larger \textit{N} when one includes decoherence or detection inefficiencies. In other words, the all-versus-nothing behavior that seemed attractive at extreme delocalizations $N \gg 2$ is quickly offset by a degradation of nonlocality under realistic conditions.

The outline of this article is as follows. We describe the W state in Sec. \ref{sec:TheAttenuatedWState} and derive its diluted form which will be needed to simulate experimental nuisances such as losses and detection inefficiencies. We then recapitulate the Bell inequality and its associated measurements in Sec. \ref{sec:TheBellInequality}. The measurement process is formalized in Sec. \ref{sec:OptimalPOVM} with the use of of optimal positive operator-valued measures (POVM). The operators we present are optimal in the sense that, notwithstanding losses and detection inefficiencies, they perform the required projectors deterministically and accurately. Virtually no laboratory device can project optimally, but we expressly choose this best-case scenario in our derivations to prove that the all-versus-nothing behavior hoped-for in \cite{Heaney} is not possible with any conceivable measurement device. The results of our simulations are summarized in Sec. \ref{sec:Results} where we also briefly present what happens when one uses hybrid Bell measurements involving photon counting and quadrature binning.

% Realistic Bell test
%%%%%%%%%%%%%%%%%%%%%%%%%%%%%%%%%%%%%%%%%%%%%%%%%%%%%%%%%%%%%%%%%%%%%
\section{Realistic Bell test}
\label{sec:RealisticBellTest}

% The attenuated W state
%%%%%%%%%%%%%%%%%%%%%%%%%%%%%%%%%%%%%%%%%%%%
\subsection{The attenuated W state}
\label{sec:TheAttenuatedWState}

The W state consists of a single particle that is symmetrically distributed over $N$ modes. In its pure form, it reads
\beq
\ket{W} = \frac{1}{\sqrt{N}} \sum_{k=1}^N \ket{\gamma_k}, \label{eq:pureWket}
\eeq
where $\ket{\gamma_k} \equiv \ket{0}^{\otimes k-1}\otimes\ket{1}\otimes\ket{0}^{\otimes N-k}$ represents a photon at the $k$th mode, all other modes remaining empty. It is written in matrix notation as
\beq
\rhohat_{W} = \frac{1}{N} \sum_{i = 1}^N \sum_{j = 1}^N \ketbra{\gamma_i}{\gamma_j}. \label{eq:pureW}
\eeq

The above is a pure state which will inevitably suffer decoherence under realistic conditions. Not only will it
 undergo mixing and losses in preparation and transmission, but its very characterization will also incur detection inefficiencies. All these ``real-world'' effects can be bundled together in a generic decoherence factor at each of the modes. We do this by simulating a fictitious beam splitter of transmission $\eta_k^2$ at each mode $k$ of the $N$-mode system. $\eta_k^2$ can thus be interpreted as both the transmission efficiency of mode $k$ and the quantum efficiency of any measurement that is ultimately performed on it. The initially pure W state (\ref{eq:pureW}) effectively turns into the mixed state
\beqa
\rhohat_{W}' & = & \frac{1}{N} \sum_{i=1}^N \tes{\eta_i^2 \ketbra{\gamma_i}{\gamma_i} + \tes{1-\eta_i^2} \ket{0}^{\otimes N} \bra{0}^{\otimes N}} + \nonumber\\
& & \frac{1}{N} \sum_{i=1}^N \eta_i \sum_{j=i+1}^N \eta_j \tes{\ketbra{\gamma_i}{\gamma_j} + \ketbra{\gamma_j}{\gamma_i}}.
\label{eq:attenuatedW}
\eeqa

\begin{figure*}[ht]
\includegraphics{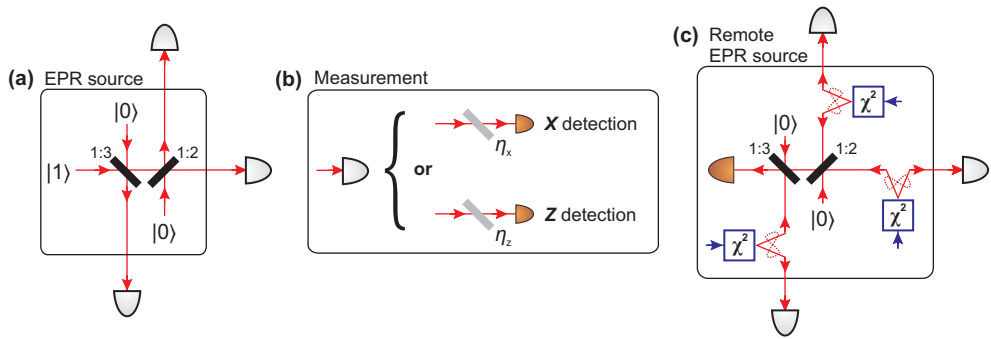}
\caption{\textbf{Experimental setup for the tripartite example (\textit{N} = 3). a,} Conceptual setup for the preparation of the W state.  A single photon is fed into two consecutive beam splitters of reflectivities $\frac{1}{3}$ and $\frac{1}{2}$, respectively. The generation of the input single photon can be achieved by heralding one photon from the pair emitted via spontaneous parametric down-conversion (SPDC). Any inefficiency of the heralding detector will not affect the purity of the produced state, but will simply reduce its generation rate. \textbf{b,} Measurement scheme. Each of the three modes that emerge are randomly measured in either one of two ways: By a projection on $\Zhat$, or by a projection on $\Xhat$. Fictitious beam splitters, drawn here in gray, are merely used as mathematical models for inefficient detection. If one works with optimal detectors, then the effectively measured state is given by (\ref{eq:attenuatedW}). \textbf{c,} Remote preparation of the W state. Three weak squeezers (labeled $\chi^2$) produce a state $p_0 \ket{00} + p_1 \ket{11}$ (with $p_0 \gg p_1$) where one mode from each is sent ``backwards'' to the same beam splitter arrangement as in (\textbf{a}) for entanglement with the other two. Upon detection of a single photon, the remote modes collapse to a W state.}
\label{fig:setup}
\end{figure*}

From an experimental point of view, it is worth highlighting the relative ease with which W states can be prepared in comparison to, say, GHZ states. It indeed suffices to send a single photon into two beam splitters with reflectivities of $\frac{1}{3}$ and $\frac{1}{2}$, respectively (Fig. \ref{fig:setup}a). A remote preparation can also be implemented in order to avoid transmitting the W state through lossy channels (Fig. \ref{fig:setup}c) \cite{RemotePrep}.

% The Bell inequality
%%%%%%%%%%%%%%%%%%%%%%%%%%%%%%%%%%%%%%%%%%%%
\subsection{The Bell inequality}
\label{sec:TheBellInequality}

The Bell factor $\Omega$ pertaining to W states was initially derived in \cite{Cabello} and extended to the $N$-partite case in \cite{Heaney}. Any model that satisfies the inequality $\Omega > 0$ has a probability $\Omega$ of contradicting local realism. For a permutationally-symmetric state such as the W, a closed-form expression for the Bell factor is given by
\beqa
\Omega & = & N \cdot P(z_1 = -1, z_2 = \ldots = z_N = +1) \nonumber\\
& & - (N)_2 \cdot P(x_1 = +1, x_2 = -1, z_3 = \ldots = z_N = +1) \nonumber\\
& & - P(x_1 = \ldots = x_N = +1) \nonumber\\
& & - P(x_1 = \ldots = x_N = -1) \label{eq:Omega}
\eeqa
where $(N)_2 = N(N-1)$ is the number of permutations of any pair of modes over a total of $N$. The function $P$ is the probability of projecting the W state on a set of Pauli operators $\Xhat$ or $\Zhat$ so as to achieve the eigenvalues specified as arguments. The subscript on the eigenvalues labels the measured mode. In the Fock basis, the eigenvalue equations of the $\Xhat$ and $\Zhat$ operators are
\beq
\left\{
	\begin{array}{lr}
		\Zhat\ket{0} = +1\ket{0}, \Zhat\ket{1} = -1\ket{1}, \mbox{ and } \\
		\Xhat\hak{\frac{1}{\sqrt{2}}\tes{\ket{0}\pm\ket{1}}} = \pm\frac{1}{\sqrt{2}}\tes{\ket{0}\pm\ket{1}}.
	\end{array}
\right. \label{eq:Projectors}
\eeq
We shall return to the formal representation of these operators in the next section. For now, it can be seen that $\Zhat$ is essentially a binary photon detector which takes on eigenvalue +1 or $-1$ depending on whether a photon is measured. $\Xhat$ is not an operator in the energy basis, it could however be aligned with the Fock basis via a Hadamard rotation as proposed in \cite{Heaney}. That said, we shall abstract these operators from their physical implementation. Instead, we assume that optimal POVM's exist for $\Zhat$ and $\Xhat$ and that one could in principle perform unambiguous projections reproducing (\ref{eq:Projectors}). Recall, however, that our use of optimal POVM's does not dispense us from detection inefficiencies; these have already be taken into account in the derivation of the diluted state $\rhohat_{W}'$ above.

% Optimal POVM
%%%%%%%%%%%%%%%%%%%%%%%%%%%%%%%%%%%%%%%%%%%%%%%%%%%%%%%%%%%%%%%%%%%%%
\section{Optimal POVM}
\label{sec:OptimalPOVM}

Consider a generic qubit consisting of an equal superposition of a single photon and the vacuum
\beq
\ket{\psi_{\theta,\phi}} = \cos(\theta)\ket{0}+\sin(\theta)e^{i\phi}\ket{1},
\eeq
where $\theta$ and $\phi$ are, respectively, the azimuthal and equatorial coordinates on the Bloch sphere. From now on, we shall ignore the equatorial dimension and only work on the circle spanned by $\theta$. The POVM which optimally projects on $\ket{\psi_{\theta}}$ is represented in the Fock basis as
\beq
\Pihat_{\theta} = \ketbra{\psi_{\theta}}{\psi_{\theta}} = \begin{bmatrix}
  \cos^2(\theta) & \cos(\theta)\sin(\theta)\\
  \sin(\theta)\cos(\theta) & \sin^2(\theta)\\
\end{bmatrix}.
\eeq
The projective probability $P_{q,\theta}$ of any qubit $\rhohat_{q}$ on $\Pihat_{\theta}$ is therefore given by
\beq
P_{q,\theta} = \mbox{Tr}\braces{\Pihat_{\theta}\cdot\rhohat_{q}}. \label{eq:ProjectiveProbability}
\eeq

Going back to the two projections of interest to us, the measurement operators satisfying the eigenvalue equations (\ref{eq:Projectors}) are
\beqa
\Pihat_{z}^{+} & = & \Pihat_{0} = \begin{bmatrix}
  1 & 0 \\
  0 & 0 \\
\end{bmatrix} \mbox{, } \Pihat_{z}^{-} = \Pihat_{\frac{\pi}{2}} = \begin{bmatrix}
  0 & 0 \\
  0 & 1 \\
\end{bmatrix} \mbox{, and}
\nonumber\\
\Pihat_{x}^{+} & = & \Pihat_{\frac{\pi}{4}} = \begin{bmatrix}
  \frac{1}{2} & \frac{1}{2} \\
  \frac{1}{2} & \frac{1}{2} \\
\end{bmatrix} \mbox{, } \Pihat_{x}^{-} = \Pihat_{\frac{7\pi}{4}} = \begin{bmatrix}
  \frac{1}{2} & \mbox{-}\frac{1}{2} \\
  \mbox{-}\frac{1}{2} & \frac{1}{2} \\
\end{bmatrix},
\eeqa 
where the superscript on $\Pihat$ indicates the sign of the eigenvalue.

% Results and discussion
%%%%%%%%%%%%%%%%%%%%%%%%%%%%%%%%%%%%%%%%%%%%%%%%%%%%%%%%%%%%%%%%%%%%%
\section{Results and discussion}
\label{sec:Results}

% Best-case scenario
%%%%%%%%%%%%%%%%%%%%%%%%%%%%%%%%%%%%%%%%%%%%
\subsection{Best-case scenario}

With the POVM's for $\Zhat$ and $\Xhat$ projections at our disposal, the computation of Bell's factor (\ref{eq:Omega}) is given by its constituent probabilities, which are themselves $N$-mode extensions of (\ref{eq:ProjectiveProbability}):
\begin{widetext}
\beqa
P(z_1 = -1, z_2 = \cdots = z_N = +1) & = & \mbox{Tr}\braces{\hak{\Pihat_z^{-} \otimes \tes{\Pihat_z^{+}}^{\otimes N-1}} \cdot \rhohat_{W}'} \label{eq:P1},\\
P(x_1 = +1, x_2 = -1, z_3 = \cdots = z_N = +1) & = & \mbox{Tr}\braces{\hak{\Pihat_x^{+} \otimes \Pihat_x^{-} \otimes \tes{\Pihat_z^{+}}^{\otimes N-2}} \cdot \rhohat_{W}'}, \label{eq:P2}\\
P(x_1 = \cdots = x_N = \pm 1) & = & \mbox{Tr}\braces{\hak{\Pihat_x^{\pm}}^{\otimes N} \cdot \rhohat_{W}'}. \label{eq:P3}
\eeqa
\end{widetext}

\begin{figure}[ht]
\includegraphics[scale=.84]{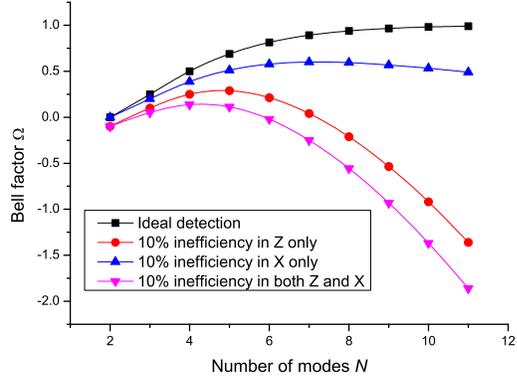} % The actual EPS is 1200 pixels wide
\caption{(Color online). Scaling of the Bell factor with the number $N$ of delocalized modes. Four scenarios are considered regarding the detection inefficiencies: ideal detection (squares), 10\% detection inefficiency in $\Zhat$ (circles), 10\% detection inefficiency in $\Xhat$ (upward triangles), 10\% detection inefficiency in both $\Zhat$ and $\Xhat$ (downward triangles). It is worth noting that inefficiencies in $\Zhat$ are more detrimental to the Bell factor than inefficiencies in $\Xhat$.}
\label{fig:Omega_vs_N}
\end{figure}

Evaluating the above probabilities yields the analytical expression
\beqa
\Omega(\eta_{z}, \eta_{x}) & = & \frac{\eta_{z}^2}{2}\tes{3+\frac{N^2}{2}-\frac{3}{2}N} - 2^{1-N} - \frac{N^2}{4} + \frac{N}{4} \nonumber\\
& & + \eta_{x}^2 \tes{\frac{N}{2} + 2^{1-N} - 2^{1-N}N - \frac{1}{2}}, \label{eq:OmegaClosedForm}
\eeqa
where $\eta_{z}^2$ and $\eta_{x}^2$ are the efficiencies of the $\Zhat$ and $\Xhat$ measurements, respectively.

In the ideal case, the Bell factor reduces to
\beq
\Omega(\eta_{z}=\eta_{x}=0) = 1-\frac{N}{2^{N-1}}. \label{eq:OmegaLossless}
\eeq
It is this expression (\ref{eq:OmegaLossless}) which prompted the optimism of Heaney \textit{et al.} \cite{Heaney} towards the W state and its potential to exhibit robust violations of locality. One can indeed see that $\lim_{N \to +\infty} \Omega = 1$, implying an essentially all-versus-nothing behavior of the W at large $N$. However, as soon as one brings in non-unity detection efficiencies, the scaling of the Bell factor with larger $N$ eventually curves downwards to ultimately dip beneath the locality bound. Four sample trends of the Bell factor as a function of the number of modes $N$ are shown in Fig. \ref{fig:Omega_vs_N}. 

\begin{figure}[ht]
\includegraphics[scale=.84]{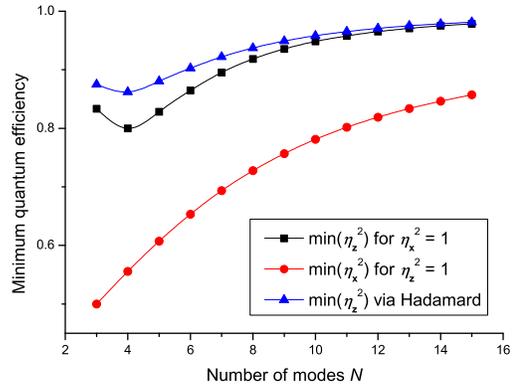}
\caption{(Color online) Minimum quantum efficiency for either $\Zhat$ or $\Xhat$ as a function of the number of modes. These plots are obtained by solving $\Omega = 0$ in (\ref{eq:OmegaClosedForm}) for $\eta_z^2$ (squares) and $\eta_x^2$ (circles) while maintaining $\eta_x^2 = 1$ and $\eta_z^2 = 1$, respectively. The top curve traces the minimum quantum efficiency required of photon detectors if one is to use a Hadamard rotation to perform $\Xhat$ measurements from $\Zhat$ basis (triangles).}
\label{fig:minEfficiency_vs_N}
\end{figure}

A further insight into the scaling of the Bell factor with the number of modes $N$ is obtained by looking at the minimum quantum efficiencies required if any violation of locality is to be witnessed. Fig. \ref{fig:minEfficiency_vs_N} shows the trend in min($\eta_z^2$) and min($\eta_x^2$) for $\eta_x^2 = 1$ and $\eta_z^2 = 1$, respectively. The key result is that any quantum efficiency of less than either 80\% for $\Zhat$ detection (assuming perfect $\Xhat$) or 50\% for $\Xhat$ detection (assuming perfect $\Zhat$) will prohibit any display of nonlocality. 

If, as suggested by Heaney \textit{et al.} \cite{Heaney}, the $\Xhat$ projection is achieved by an ideal Hadamard rotation followed by a $\Zhat$ detection, then the effect of $\eta_z^2$ is felt on all \textit{N} measurement sites. This is drawn as the upper curve on Fig. \ref{fig:minEfficiency_vs_N}. The minimum quantum efficiency thus required by the scheme of Heaney \textit{et al.} is $\eta_{z}^2 = 86.2\%$ for $N = 4$. Any scaling to larger $N$ will not help in decreasing the minimum quantum efficiency.

% Realistic scenario: Hybrid measurements
%%%%%%%%%%%%%%%%%%%%%%%%%%%%%%%%%%%%%%%%%%%%
\subsection{Realistic scenario: Hybrid measurements}
\label{sec:HybridMeasurements}

It may be worthwhile at this point to use measurement projectors for which there actually exists laboratory devices. One obvious choice for $\Zhat$ is the avalanche photodiode (APD) which, ignoring dark counts and inefficiencies, clicks when at least one photon is detected. For the $\Xhat$ operation, whose Hamiltonian is not diagonal in the energy eigenbasis, one can resort to binning continuous variables acquired by a homodyne measurement. This hybrid method of detection has been propounded recently in \cite{Cavalcanti} in the context of N00N states but could conceivably be reused for other systems thanks to the high efficiency offered by homodyning. 

\begin{figure}[ht]
\includegraphics[scale=1.1]{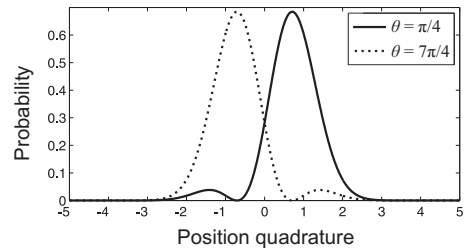}
\caption{Quadrature probability distribution of the qubit $\cos(\theta)\ket{0}+\sin(\theta)\ket{1}$ for $\theta = \frac{\pi}{4}$ and $\theta = \frac{7\pi}{4}$.}
\label{fig:quadratures}
\end{figure}

Let's look at how quadrature-binning can be used to implement an $\Xhat$-measurement. The binning is motivated by the symmetry of the quadrature probability distribution of $\ket{\psi_{\frac{\pi}{4}}}$ (i.e., $x=+1$) and $\ket{\psi_{\frac{7\pi}{4}}}$ (i.e., $x=-1$) about the zero in-phase quadrature (Fig. \ref{fig:quadratures}). We shall therefore bin our quadrature measurements by assigning $q < 0$ to $x = -1$ and $q > 0$ to $x = +1$. (Note that an ambiguity in this measurement process will arise from the overlap of the quadrature distributions around $q = 0$.) In the Fock representation, $\Xhat$ projections based on homodyne binning are given by
\beq
\Pihat_{\theta}^{\mbox{\tiny HD}} = \begin{bmatrix}
  \chi_{0,0} & \chi_{0,1} & \cdots & \chi_{0, \mbox{\tiny M}} \\
  \chi_{1,0} & \chi_{1,1} & \cdots & \vdots \\
  \vdots & \vdots & \ddots & \vdots \\  
	\chi_{\mbox{\tiny M},0} & \chi_{\mbox{\tiny M},1} & \cdots & \chi_{\mbox{\tiny M}, \mbox{\tiny M}} \\
\end{bmatrix}, \label{eq:PIdiscrete}
\eeq
where $\chi_{n,m} = \int_{\Delta(\theta)} \! \overline{\phi_{m}(q)}\phi_{n}(q) \, \mathrm{d}q$ and $\phi_{k}(q) = \braket{q}{k}$ is the quadrature probability distribution of a Fock state $\ket{n}$. $M$ is the maximum number of photons inhabiting the measured mode and is set to $M = 1$ in the particular case of single-photon qubits we are dealing with. Note the dependence of the post-selection range $\Delta(\theta)$ on the witness qubit angle $\theta$. In analogy to (\ref{eq:ProjectiveProbability}), the projective probabilities for a matrix element $\ketbra{n}{m}$ on the two possible eigenvectors of $\Xhat$ are
\beqa
P_{\mbox{\scriptsize $\ketbra{n}{m},\frac{7\pi}{4}$}} & = & \int_{-\infty}^{0} \overline{\phi_{m}(q)}\phi_{n}(q) \, \mathrm{d}q, \label{eqna:Pxn} \\
P_{\mbox{\scriptsize $\ketbra{n}{m},\frac{\pi}{4}$}} & = & \int_{0}^{\infty} \overline{\phi_{m}(q)}\phi_{n}(q)  \, \mathrm{d}q. \label{eqna:Pxp}
\eeqa

With these projective probabilities in hand, an evaluation of the Bell factor via Eqs. (\ref{eq:P1})-(\ref{eq:P3}) yields
\beqa
\Omega_{\mbox{\scriptsize hybrid}}(\eta_{\mbox{\tiny APD}}, \eta_{\mbox{\tiny HD}}) & = & \frac{\eta_{\mbox{\tiny APD}}^2}{4} \tes{N^2-3N+6} \nonumber\\
& & + \frac{\eta_{\mbox{\tiny HD}}^2}{\pi} \tes{2^{2-N} + N - 2^{2-N}N - 1}  \nonumber\\
& & + \frac{1}{4}\tes{N-N^2-2^{3-N}}, \label{eq:OmegaHybridClosedForm}
\eeqa
where $\eta_{z}^2 = \eta_{\mbox{\tiny APD}}^2$ and $\eta_{x}^2 = \eta_{\mbox{\tiny HD}}^2$ are the quantum efficiencies of the APD and homodyne detectors (HD), respectively. 

\begin{figure}[ht]
\includegraphics[scale=.84]{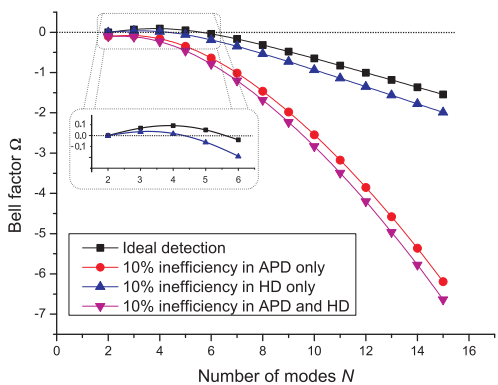}
\caption{(Color online). Scaling of the Bell factor with the number $N$ of delocalized modes in the case of hybrid detection involving an APD for $\Zhat$ and homodyne thresholding for $\Xhat$. Four scenarios are considered regarding the detection inefficiencies: ideal detection (squares), 90\% quantum efficiency for the APD (circles), 90\% detection efficiency for the homodyne detector (upward triangles), and 90\% detection efficiency for both detectors (downward triangles).}
\label{fig:Omega_vs_N_hybrid}
\end{figure}

Fig. \ref{fig:Omega_vs_N_hybrid} shows the scaling of the Bell factor with \textit{N}. One predictable observation is that the experimental evaluation of the Bell factor with this hybrid scheme leads to much smaller violation margins than those obtained by optimal POVM's. Of particular relevance to the present article is the fact that scaling to larger \textit{N} is not monotonic: Even with unit efficiencies, the Bell factor barely skims the nonlocality bound, peaks at $\mbox{max}(\Omega_{\mbox{\scriptsize hybrid}}) \approx 0.09$ for $N = 4$ then plunges back in the locality range for $N \geq 6$. Physically, this weak violation margin is explained by the fact that (\ref{eq:PIdiscrete}) is really an approximate projector: The two orthogonal qubits $\ket{\psi_{\frac{\pi}{4}}}$ and $\ket{\psi_{\frac{7\pi}{4}}}$ cannot be perfectly resolved by homodyne measurements because of their overlapping quadrature wave functions (cf. Fig. \ref{fig:quadratures}).

\begin{figure}[ht]
\includegraphics[scale=.84]{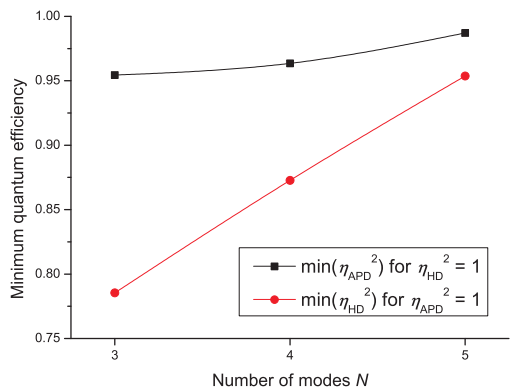}
\caption{(Color online) Minimum quantum efficiency for either the APD or the homodyne detector (HD) as a function of the number of modes. These plots are obtained by solving $\Omega_{\mbox{\scriptsize hybrid}} = 0$ in (\ref{eq:OmegaHybridClosedForm}) for $\eta_{\mbox{\tiny APD}}^2$ (squares) and $\eta_{\mbox{\tiny HD}}^2$ (circles) while maintaining $\eta_{\mbox{\tiny HD}}^2 = 1$ and $\eta_{\mbox{\tiny APD}}^2 = 1$, respectively.}
\label{fig:minEfficiency_vs_N_hybrid}
\end{figure}

The scaling of the minimum quantum efficiencies required to violate Bell's inequality are shown in Fig. \ref{fig:minEfficiency_vs_N_hybrid}. The most salient result is that nonlocality cannot be shown by the hybrid measurement scheme described above for any system with $N \geq 6$. Indeed, the ``minimum quantum efficiencies'' beyond $N = 5$ take on unphysical values above unity. The increase of the violation margin with larger $N$ has therefore been overwhelmingly offset by a decrease in purity. The best result that can be achieved is for $N = 3$ where the minimum quantum efficiencies required are 95\% for the APD (assuming ideal homodyning) or 79\% for the the homodyne detector (assuming ideal photon detection).

% Conclusion
%%%%%%%%%%%%%%%%%%%%%%%%%%%%%%%%%%%%%%%%%%%%%%%%%%%%%%%%%%%%%%%%%%%%%
\section{Conclusion}
\label{sec:Conclusion}

We have shown that in the context of demonstrating nonlocality with W states, the transition from theory to experiment is not only quantitative, but also qualitative. The non-statistical violation of local realism which could effectively have been achieved by W states for large \textit{N} turns out to be invalidated by the slightest introduction of decoherence, even if the projective measurements were optimal. In practice, one would have to make the unrealistic assumption that deterministic Hadamard rotations can be used on Fock states superpositions. Even then, the minimum quantum efficiency for photon detection would have to be 86.2\%. Such high requirements can only be met by demanding detection setups (e.g., transition-edge sensors \cite{Rosenberg}). The burden on quantum efficiency has therefore been shown not to benefit in any appreciable way from large delocalizations of a single quantum. The situation is predictably worse when one uses approximate projectors to perform the Pauli projector $\Xhat$: A minimum quantum efficiency of 95\% is required of the photon counters if the Bell measurements are performed with a hybrid scheme involving quadrature binning.

\textit{Note}: It has come to our attention that similar work has very recently been published by Chaves and Bohr Brask in \cite{Chaves2011}.

% Bibliography
%%%%%%%%%%%%%%%%%%%%%%%%%%%%%%%%%%%%%%%%%%%%%%%%%%%%%%%%%%%%%%%%%%%%%

\end{document}